\documentclass{article}

\PassOptionsToPackage{numbers, compress}{natbib}
\usepackage[preprint]{neurips_2023}

\usepackage[utf8]{inputenc} 
\usepackage[T1]{fontenc}    
\usepackage{hyperref}       
\usepackage{url}            
\usepackage{booktabs}       
\usepackage{amsfonts}       
\usepackage{nicefrac}       
\usepackage{microtype}      
\usepackage{xcolor}         
\usepackage{amsmath}
\usepackage{amssymb}
\usepackage{mathtools}
\usepackage{amsthm}
\usepackage{multirow}
\usepackage{wrapfig}
\usepackage[capitalize,noabbrev]{cleveref}
\usepackage{graphicx}
\usepackage{subfigure} 
\usepackage{float}
\usepackage{algorithm} 
\usepackage{algorithmic}
\usepackage{textcomp} 
\usepackage{listings} 
\usepackage{enumitem}
\usepackage{utfsym}
\usepackage{times}
\usepackage{latexsym}
\usepackage{fancyvrb}
\usepackage{CJKutf8}

\newcommand{\commentout}[1]{}

\title{FireRedTTS-2: Towards Long Conversational Speech Generation for Podcast and Chatbot}

\author{
Kun Xie$^*$, 
Feiyu Shen\thanks{Equal contribution} ,
Junjie Li, 
Fenglong Xie\thanks{Corresponding Author: Fenglong Xie (fenglongxie@xiaohongshu.com)} ,
Xu Tang, 
Yao Hu\\
Xiaohongshu \\
}

\begin{document}
\maketitle

\begin{abstract}
Current dialogue generation approaches typically require the complete dialogue text before synthesis and produce a single, inseparable speech containing all voices, making them unsuitable for interactive chat; moreover, they suffer from unstable synthesis, inaccurate speaker transitions, and incoherent prosody.
In this work, we present FireRedTTS‑2, a long-form streaming TTS system for multi-speaker dialogue generation, delivering stable, natural speech with reliable speaker switching and context-aware prosody. A new 12.5Hz streaming speech tokenizer accelerates training and inference, extends maximum dialogue length, encodes richer semantics to stabilize text-to-token modeling and supports high-fidelity streaming generation for real-time applications. We adopt a text–speech interleaved format, concatenating speaker-labeled text with aligned speech tokens in chronological order, and model it with a dual-transformer: a large decoder-only transformer predicts tokens at the first layer, and a smaller one completes subsequent layers. Experimental results show that FireRedTTS‑2 integrates seamlessly with chat frameworks and, with minimal fine-tuning, produces emotionally expressive speech guided by implicit contextual cues. 
In podcast generation, it surpasses existing systems including MoonCast, Zipvoice-Dialogue, and MOSS-TTSD in objective intelligibility, speaker-turn reliability, and perceived naturalness with context-consistent prosody.
Our demos are available at \url{https://fireredteam.github.io/demos/firered_tts_2}. 
\end{abstract}

\section{Introduction}
\label{sec:intro}
Large language model (LLM) based text-to-speech (TTS) systems can generate natural-sounding speech with zero-shot voice cloning and are widely used for monologue applications like video dubbing. These systems typically follow one of two modeling paradigms: an autoregressive, decoder-only transformer that predicts speech tokens\cite{guo2024fireredtts1, guo2025fireredtts1s, du2024cosyvoice1, du2024cosyvoice2, du2025cosyvoice3, deng2025indextts, wang2025sparktts}, or a non-autoregressive flow-matching model that produces mel-spectrograms directly from text\cite{chen2024f5tts, eskimez2024e2tts}. While these monologue TTS systems can be adapted to dialogue generation by segmenting dialogue text and synthesizing each fragment independently\cite{huang2025stepaudio, huang2023audiogpt, xiao2025podagent}, this strategy ignores preceding text and speech context, leading to a loss of conversational coherence.

Recent works have extended TTS system to two-speaker dialogue generation, which can be grouped into three categories based on how text and speech are organized across turns: (1) splitting the dialogue text into two parallel channels and synthesizing a single mixed speech track containing both voices, which can naturally handles overlapping speech and generate interjections effects\cite{zhang2024covomix1, zhang2025covomix2}; (2) concatenating the dialogue text in chronological order with each utterance prefixed by a speaker label, which likewise produces a mixed speech track\cite{ju2025mooncast, nari2025dia, darefsky2024parakeet, zhu2025zipvoice, moss2025ttsd, peng2025vibevoice}; and (3) interleaving the text and speech of each utterance\cite{Schalkwyk2025sesame}. Approaches (1) and (2) require the complete dialogue text before synthesis and yield a single inseparable mixed speech, limiting their suitability for interactive scenarios such as chat, whereas (3) supports flexible sentence-by-sentence generation, suitable for both interactive chat and podcast production.

In this work, we present FireRedTTS‑2, a long‑form, streaming TTS system for multi‑speaker dialogue and podcast generation that delivers stable, natural speech, reliable speaker switching, and context‑aware prosody. A new streaming 12.5Hz speech tokenizer accelerates training and inference, lengthens the effective dialogue context, encodes richer semantics to stabilize text‑to‑token modeling and supports high-fidelity streaming generation for real-time applications. We adopt an interleaved text–speech format by concatenating speaker‑labeled text with speech tokens in chronological order, and model it with a dual‑transformer architecture: a large decoder‑only network predicts tokens at the first layer, while a smaller network refines the subsequent layers. Experimental results show that FireRedTTS‑2 integrates seamlessly with chat frameworks and, with minimal fine‑tuning, produces emotionally expressive speech guided by implicit context. In podcast generation, it surpasses the state of the art systems including MoonCast\cite{ju2025mooncast}, ZipVoice-Dialogue\cite{zhu2025zipvoice}, and MOSS-TTSD\cite{moss2025ttsd} in objective intelligibility, speaker‑turn reliability, and perceived naturalness, while maintaining prosody consistent with long‑range context.

\section{FireRedTTS-2}
\begin{figure}[htp]
\centering
\includegraphics[width=\linewidth]{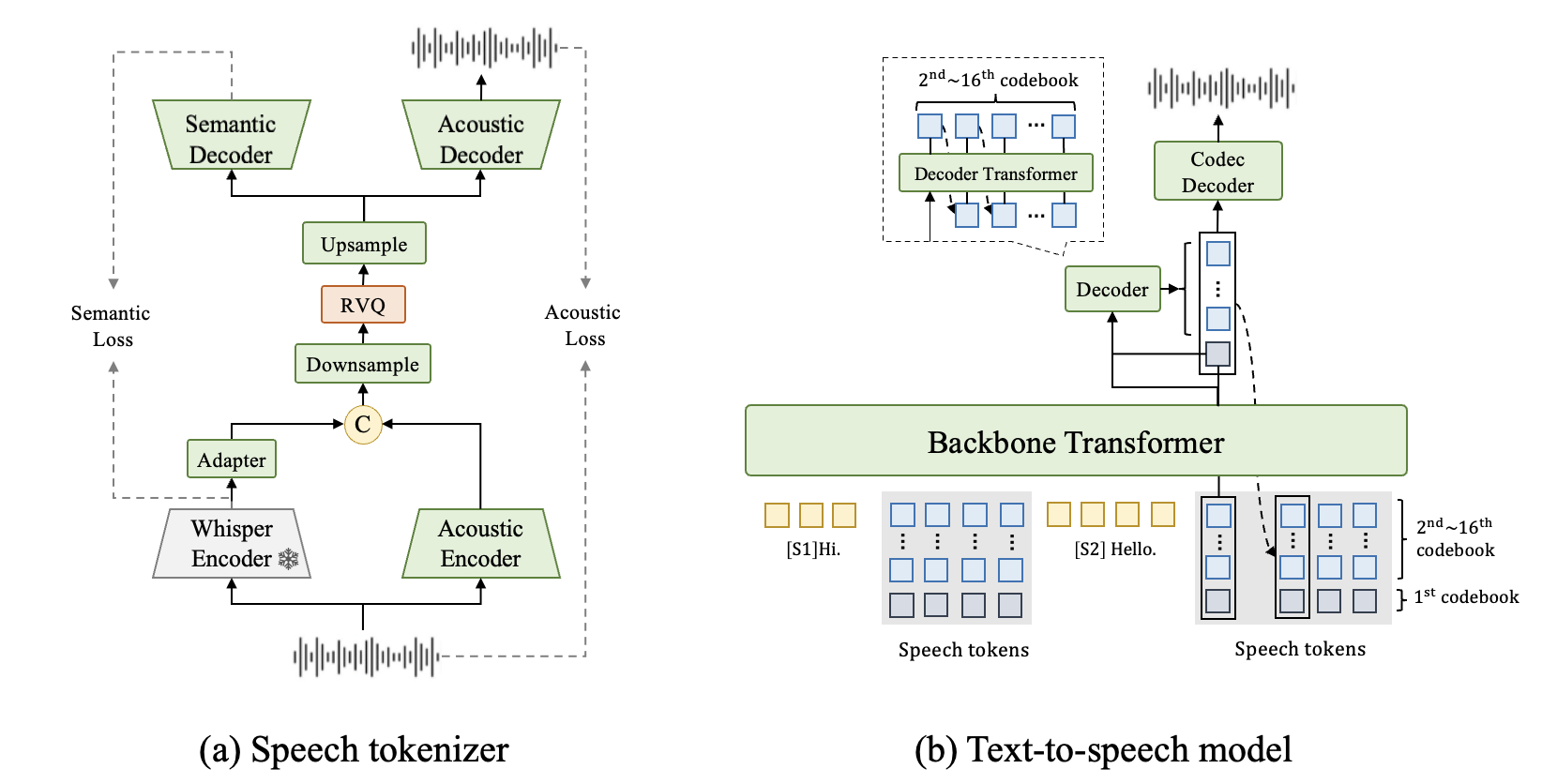}
\caption{An overview of FireRedTTS-2, including: (a) a new speech tokenizer with a 12.5Hz frame rate and enhanced semantic information, and (b) a text-to-speech model using a dual-transformer architecture with interleaved text–speech input, enabling sentence-by-sentence generation and contextually coherent prosody.}
\label{img:tts_model_framework}
\end{figure}

As shown in Figure \ref{img:tts_model_framework}, FireRedTTS-2 consists of a newly developed speech tokenizer and a text-to-speech model with perception to previous text and speech context.  

\subsection{Speech Tokenizer}
We design our speech tokenizer to enhance dialogue modeling, with a focus on long, multi-speaker speech sequence. To make such sequences tractable, we reduce the frame rate to 12.5Hz, half that of most open-source tokenizers\cite{du2024cosyvoice1, du2024cosyvoice2, du2025cosyvoice3, ye2025xcodec1, ye2025llasa, zhang2023speechtokenizer}. We further employ semantic injection and supervision to simplify text-to-token modeling, which has been shown to improve synthesis stability\cite{moss2025ttsd, ye2025xcodec1, ye2025llasa, zhang2023speechtokenizer, defossez2024moshi}. In addition, the tokenizer supports high-fidelity streaming generation for real-time applications.

As illustrated in Figure \ref{img:tts_model_framework}(a), our speech tokenizer employs a pretrained Whisper\cite{radford2023whisper} encoder to extract semantic features from the 16kHz input speech. These semantic features are encoded by an adapter and then concatenated with acoustic features from a trainable acoustic encoder structurally identical to the Whisper encoder. The combined features undergo 4 times downsampling from 50Hz to 12.5Hz and are discretized by a residual vector quantizer (RVQ)\cite{zeghidour2021soundstream} with 16 layers, each containing 2048 code entries. The quantized features are upsampled to 50Hz and fed to a semantic decoder to predict the original semantic features derived from the pretrained Whisper encoder. The same upsampled features are also used by a Vocos\cite{siuzdak2023vocos}-based acoustic decoder to reconstruct the waveform. Depending on the reception fields of its inner convolution and attention layers, the acoustic decoder can be implemented as either streaming or non-streaming.


To balance generalization capability and speech quality, we train our speech tokenizer in two stages similar to\cite{moss2025ttsd}. First, the acoustic decoder is implemented as non-streaming and optimized to predict 16kHz speech. We use approximately 500k hours of speech data and train the model for 320k steps on 32 H800 GPUs, with each sample randomly cropped to 6 seconds. For the final 35k steps, we incorporate the perceptual loss\cite{ye2025llasa, parker2024scaling} to further improve semantic details. In the second stage, we freeze the encoding part and replace the acoustic decoder with a fully streaming variant that predicts 24kHz speech. We continue to train the speech tokenizer on a subset of 60k hours high-fidelity speech data for 80k steps.

\subsection{Text-to-Speech Model}
Building on the new speech tokenizer, we employ a dual-transformer architecture akin to \cite{Schalkwyk2025sesame,defossez2024moshi} that operates on a text–speech interleaved sequence, enabling flexible sentence-by-sentence generation and reducing first-packet latency. As illustrated in Figure \ref{img:tts_model_framework}(b), each dialogue text is prefixed with a speaker tag (e.g., "[S1]") and concatenated with its corresponding speech tokens; these segments are then joined in temporal order to form sequences such as "[S1]<text><audio>[S2]<text><audio>[S3]<text><audio>...". 
Existing approaches\cite{nari2025dia, darefsky2024parakeet, moss2025ttsd} model multi-layer speech tokens using the delay-pattern\cite{copet2023musicgen}: for $N$ token layers, the $i^\text{th}$ layer is shifted $i-1$ timesteps to the right, and $N$ prediction heads predict these shifted layers in parallel. This design has two main drawbacks: first, at each timestep the model has only partial access to the speech tokens from previous steps due to the rightward shifts, weakening contextual conditioning; second, obtaining the complete set of $N$ layer tokens for the first timestep requires $N$ autoregressive steps, resulting in high latency.
To overcome these issues, we adopt a dual-transformer architecture comprising a backbone transformer that processes the text–speech interleaved sequence and predicts the first-layer tokens, and a smaller decoder transformer that generates remaining token layers. Both transformers are based on Qwen2.5\cite{ahmed2025qwen25} structure. At each timestep, the decoder consumes both the predicted first layer token and the backbone’s hidden states, which provide complete contextual information. Comparing with the delay-pattern, it requires one auto-regressive inference step of the backbone transformer and $N-1$ steps of the smaller decoder, reducing computation and first-packet latency. Moreover, our speech tokenizer produces high-fidelity speech in a streaming manner without requiring separate token-to-speech modules, simplifying the overall system.

The text-to-speech model is optimized with the following loss function:
\begin{align}
\mathcal{L}_{loss}=2*((1-\lambda_{decoder})\mathcal{L}_{backbone}+\lambda_{decoder}\mathcal{L}_{decoder})+\lambda_{text}\mathcal{L}_{text}
\end{align}
Here, $\mathcal{L}_{backbone}$ and $\mathcal{L}_{decoder}$ denote the cross-entropy loss of the backbone and decoder transformer respectively. To improve training efficiency, we optimize the decoder transformer only on 1/8 of the speech segments in the interleaved sequence. Additionally, we incorporate a cross-entropy loss for the textual part ($\mathcal{L}_{text}$) to stabilize training. In our experiments, we set $\lambda_{text} = 0.01$ and $\lambda_{decoder} = 0.6$.

To enable the model with dialogue generation capability, we adopt a three-stage curriculum training process utilized in \cite{zhang2024covomix1, moss2025ttsd}, comprising pretraining, post-training, and supervised fine-tuning (SFT). The pretraining stage leverages 1.1M hours of monologue speech data and trains the model for 2 epochs to build foundational text-to-speech ability. Subsequently, we post-train FireRedTTS-2 for 5 epochs on 300k hours of multi-speaker dialogue data, with each dialogue containing 2 to 5 speakers, to enable robust multi-speaker dialogue generation. Finally, the SFT stage is applied to tailor the model to specific voices with minimal data.

\section{Downstream Applications}
FireRedTTS-2 excels at both monologue and dialogue speech generation. For monologues, it offers competitive zero-shot voice cloning suitable for tasks like video dubbing. For dialogues generation, it surpasses monologue TTS systems due to its perception of text and speech context, producing speech with coherent prosody. Compared with existing dialogue TTS systems, it supports sentence-by-sentence generation, enabling both interactive chat and offline podcast production. Across both modes, FireRedTTS-2 can be tailored to specific application requirements with minimal data, demonstrating strong flexibility.

\subsection{Voice Cloning}
Our speech tokenizer captures both semantic and acoustic information, enabling fine-grained acoustic modeling. Paired with large-scale pretraining on monologue speech, it allows FireRedTTS-2 to deliver robust zero-shot voice cloning.
Given a speech prompt and its transcript, we concatenate the prompt transcript, the target text to be synthesized, and the discretized prompt speech tokens, then feed this sequence into the text-to-speech model to autoregressively generate new speech tokens. The generated tokens are appended to the prompt tokens and decoded into a waveform by the speech tokenizer’s decoder, after which the portion corresponding to the original prompt is removed.

\subsection{Interactive Chat}
Current interactive chat frameworks\cite{huang2025stepaudio, huang2023audiogpt} typically rely on monologue TTS systems, which lack perception of prior user queries and system responses, often resulting in inconsistent emotion and prosody. This can be partially mitigated with explicit instructions such as emotion labels, but it requires additional fine-tuning of the text LLM and adds unnecessary complexity. 

\begin{figure}[htp]
\centering
\includegraphics[width=0.8\linewidth]{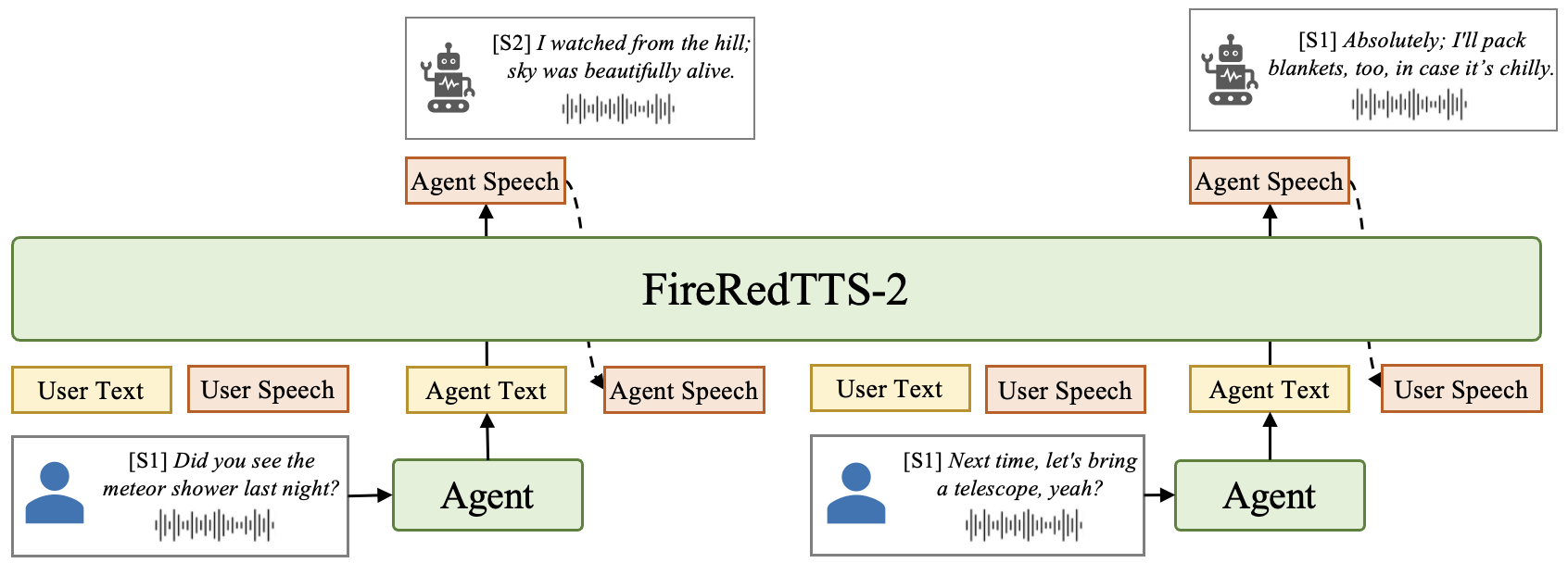}
\caption{Integration of FireRedTTS-2 into interactive chat scenarios.}
\label{img:tts2_for_chat}
\end{figure}

As shown in Figure \ref{img:tts2_for_chat}, FireRedTTS-2 can be seamlessly integrated into existing chat frameworks without modifying other modules. To address the inconsistency issue, we fine-tune the post-trained FireRedTTS-2 to infer and adjust emotion and prosody from implicit contextual cues. Specifically, we curate a 15-hour speech corpus of a distinctive female voice expressing six emotions: surprise, sadness, happiness, concern, apology, and anger. Then we emulate conversational context by first generating text context with a text LLM and then synthesizing it into speech. After fine-tuning, FireRedTTS-2 dynamically shifts emotion and tone in response to preceding chat history, delivering a near-human interactive experience.

\subsection{Podcast Generation} 
Subsequent post-training on dialogue corpus equips FireRedTTS-2 with conversational generation abilities, making it well-suited for podcast generation. Compared with conventional segmenting approaches that utlizes monologue TTS systems, it simplifies the workflow and can synthesizes contextually coherent prosody. 
Moreover, it generates dialogue speech sentence by sentence, providing greater flexibility for editing and post-processing.

\begin{figure}[htp]
\centering
\includegraphics[width=0.8\linewidth]{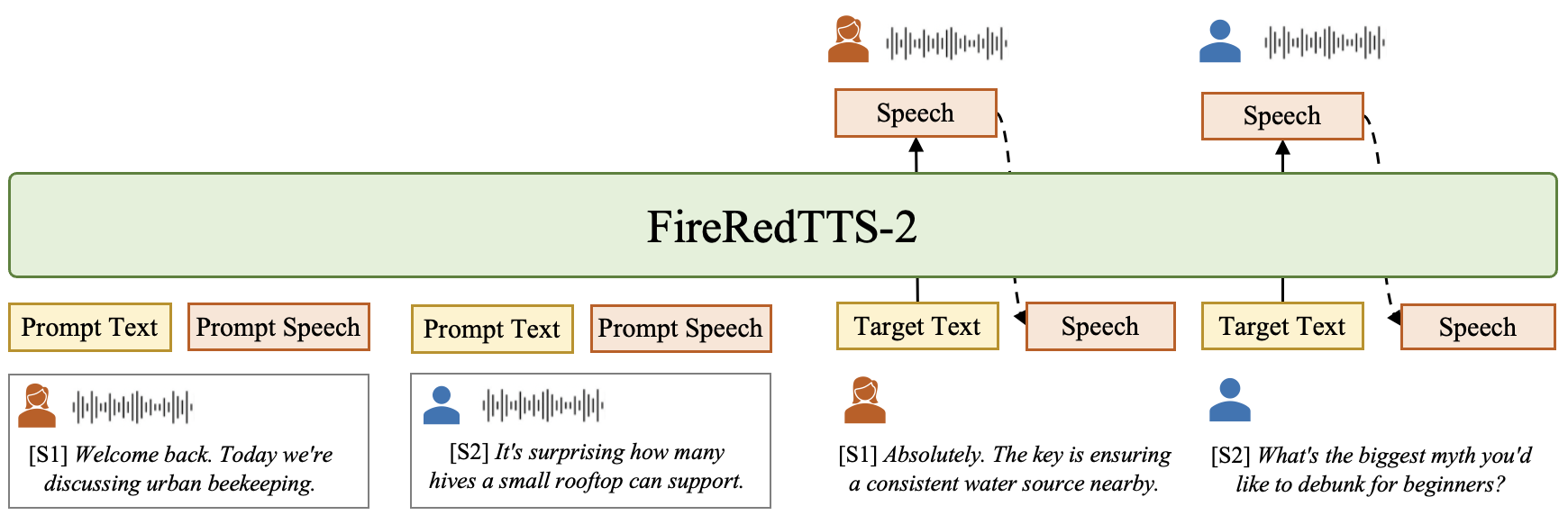}
\caption{Zero-shot podcast generation of FireRedTTS-2.}
\label{img:tts2_for_zeroshot_podcast}
\end{figure}

As shown in Figure \ref{img:tts2_for_zeroshot_podcast}, FireRedTTS-2 can perform zero-shot podcast generation by placing two dialogue turns as prompt context and then generating subsequent turns one by one. It currently supports 3 minutes dialogues with 4 speakers and can be easily scaled to longer conversations with more speakers by extending training corpus. It can also be tailored to specific speakers with minimal data. In this work, we collected approximately 50 hours of dialogue speech featuring a male and a female podcast host and fine-tuned the post-trained model for 15 epochs. Once customized,  FireRedTTS-2 delivers stable synthesis, accurate speaker transitions, and contextually coherent prosody that matches the hosts’ distinctive speaking styles.

\section{Results}

\subsection{Speech Tokenizer Evaluation}
We evaluate our speech tokenizer on intelligibility, speaker similarity, and speech quality of the reconstructed speech using the LibriSpeech test-clean set, which contains 2,620 utterances at 16 kHz. Speech intelligibility is measured by word error rate (WER) using a HuBERT-based automatic speech recognition (ASR) system\footnote{\url{https://huggingface.co/facebook/hubert-large-ls960-ft}}. Speech quality is assessed with speaker similarity (SPK-SIM) computed by the WavLM-Large model \footnote{\url{https://github.com/microsoft/UniSpeech/tree/main/downstreams/speaker_verification}}, Short-Time Objective Intelligibility (STOI), Perceptual Evaluation of Speech Quality (PESQ), and UTMOS \footnote{\url{https://github.com/tarepan/SpeechMOS}}. 
We compare our speech tokenizer with other methods that likewise incorporate semantic injection and supervision, including XY-Tokenizer\cite{moss2025ttsd}, XCodec2\cite{ye2025llasa}, SpeechTokenizer\cite{zhang2023speechtokenizer}, and Mimi\cite{defossez2024moshi}; results are reported in Table \ref{tab:codec_cmp}.

\begin{table}[htp]
\centering
\resizebox{\columnwidth}{!}{
\begin{tabular}{c|cc|cccccc}
\toprule
\textbf{Models} & \textbf{BPS} & \textbf{\begin{tabular}[c]{@{}c@{}}Frame\\ Rate\end{tabular}} & \textbf{WER$\downarrow$} & \textbf{\begin{tabular}[c]{@{}c@{}}SPK\\SIM\end{tabular}$\uparrow$} & \textbf{STOI$\uparrow$} & \textbf{\begin{tabular}[c]{@{}c@{}}PESQ\\WB\end{tabular}$\uparrow$} & \textbf{\begin{tabular}[c]{@{}c@{}}PESQ\\NB\end{tabular}$\uparrow$} & \textbf{\begin{tabular}[c]{@{}c@{}}UT\\MOS\end{tabular}$\uparrow$} \\ \midrule
Ground Truth & - & - & 1.96 & - & 1.00 & 4.64 & 4.55 & 4.09 \\ \midrule
Xcodec2 & 800 & 50 & 2.46 & 0.82 & 0.92 & 2.43 & 3.04 & \textbf{4.13} \\
XY-Tokneizer & 1000 & 12.5 & - & 0.83 & 0.91 & 2.41 & 3.00 & - \\
SpeechTokenizer & 2000 & 50 & 2.86 & 0.66 & 0.88 & 1.92 & 2.38 & 3.56 \\
Mimi & 2200 & 12.5 & 2.26 & \textbf{0.87} & \textbf{0.94} & \textbf{2.88} & \textbf{3.42} & 3.87 \\ \midrule
Ours & 2200 & 12.5 & \textbf{2.16} & \textbf{0.87} & \textbf{0.94} & 2.73 & 3.28 & 3.88 \\ \bottomrule
\end{tabular}
}
\caption{Comparison between different speech tokenizers. Best results are marked in bold.}
\label{tab:codec_cmp}
\end{table}

Table \ref{tab:codec_cmp} shows that our speech tokenizer achieves the highest intelligibility, which we attribute to the semantic injection with explicit supervision. It also ranks first or second on speaker similarity and speech quality metrics, even at a 12.5Hz frame rate, thanks to a larger quantizer that reduces quantization error and a Vocos\cite{siuzdak2023vocos}-based acoustic decoder. 
However, it trails Mimi on the PESQ metrics, likely because Mimi uses a massive, purely English training corpus that more closely matches the test set. It also scores lower on UTMOS than Xcodec2, due to its lower 12.5Hz frame rate. 
Overall, these results verify that our speech tokenizer can produce high‑quality speech despite its low 12.5Hz frame rate.

\subsection{Voice Cloning Evaluation}
We evaluate FireRedTTS-2 for voice cloning on the "Test-ZH" and "Test-EN" sets from the Seed-TTS-eval\cite{anastassiou2024seedtts} benchmark, using the model after the pretraining stage. We compare it against popular monologue TTS systems, including Seed-TTS\cite{anastassiou2024seedtts}, F5-TTS\cite{chen2024f5tts}, MaskedGCT\cite{wang2024maskgct}, SparkTTS\cite{wang2025sparktts}, and the CosyVoice series\cite{du2024cosyvoice1, du2024cosyvoice2, du2025cosyvoice3}. Results are reported in Table \ref{tab:exp_zeroshot_tts}. 

\begin{table}[htp]
\centering

\begin{tabular}{c|c|cc|cc}
\toprule
\multirow{2}{*}{\textbf{System}} & \multirow{2}{*}{\textbf{\begin{tabular}[c]{@{}c@{}}Frame\\Rate\end{tabular}}} & \multicolumn{2}{c|}{\textbf{Test-ZH}} & \multicolumn{2}{c}{\textbf{Test-EN}} \\ \cmidrule{3-6} 
& & \textbf{CER $\downarrow$} & \textbf{SIM $\uparrow$} & \textbf{WER $\downarrow$} & \textbf{SIM $\uparrow$} \\ \midrule
Human & - & 1.26 & 0.755 & 2.14 & 0.734 \\ \midrule
Seed-TTS & - & 1.12 & \textbf{0.796} & 2.25 & \textbf{0.762} \\
F5-TTS & - & 1.56 & 0.741 & \textbf{1.83} & 0.647 \\
MaskedGCT & 50 & 2.27 & 0.774 & 2.62 & 0.714 \\
SparkTTS & 50 & 1.20 & 0.672 & 1.98 & 0.584 \\ 
CosyVoice 3-1.5B & 25 & 1.12 & 0.781 & 2.21 & 0.720 \\ 
FireRedTTS-1S & 25 & \textbf{1.00} & 0.753 & 2.20 & 0.663 \\ \midrule
FireRedTTS-2 & 12.5 & 1.14 & 0.736 & 1.95 & 0.665 \\ \bottomrule
\end{tabular}
\caption{The objective evaluation on Seed-TTS test set. Best results are marked in bold.}
\label{tab:exp_zeroshot_tts}
\end{table}

Our pretrained FireRedTTS‑2 achieves 1.14\% CER on Mandarin and 1.95\% WER on English, closely matching the best results of 1.12\% CER and 1.83\% WER. We attribute this to enhanced semantic information in the speech tokens, which strengthens text‑to‑token modeling. For speaker similarity, it aligns with human recordings in Mandarin but trails in English, likely due to limited voice diversity in the English training data. Moreover, systems such as Seed-TTS inject timbre through dedicated diffusion or flow-matching modules, further boosting similarity. Compared with previous FireRedTTS‑1S, it delivers better English WER and speaker similarity, with a slight drop on Mandarin that may stem from the halved frame rate of speech tokenizer. 
However, as noted in \cite{guo2025fireredtts1s}, objective metrics cannot faithfully reflect TTS performance due to limited test set coverage and imprecise evaluation tools, and speech with more expressive prosody tends to be rated less intelligible, while plainer ones typically score higher; we therefore place greater weight on the following subjective evaluations.

\subsection{Interactive Chat Evaluation}
To assess the chat fine-tuned FireRedTTS-2’s ability to infer and adjust synthesis emotions from implicit contextual cues, we built a test set with 30 test cases for each of six emotions: surprise, sadness, happiness, concern, apology, and anger. For each test case, we use the Qwen3\cite{yang2025qwen3} model to generate a text query–response pair that implicitly conveyed the target emotion. The text query was then synthesized into speech using random voices, and FireRedTTS-2 produced the speech response. We manually label the emotion of generated speech response and calculate the emotion control accuracy. 
The results in Table \ref{tab:exp_finetune_chatbot_instruct} show that FireRedTTS-2 can infer appropriate emotions from implicit contextual cues by leveraging preceding text and speech context, thereby enabling a more human-like chat experience and validating the effectiveness of our approach.

\begin{table}[htp]
\centering
\begin{tabular}{c|cccccc}
\toprule
\multirow{2}{*}{\textbf{Model}} & \multicolumn{6}{c}{\textbf{Emotion Accuracy $\uparrow$}} \\ \cmidrule{2-7} 
 & \textbf{Surprised} & \textbf{Sad} & \textbf{Happy} & \textbf{Concern} & \textbf{Apology} & \textbf{Angry} \\ \midrule
FireRedTTS-2 & 83.3\% & 86.7\% & 90.0\% & 86.7\% & 93.3\% & 76.7\% \\ \bottomrule
\end{tabular}
\caption{Emotion control accuracy of FireRedTTS-2 after fine-tuning for interactive chat scenario.}
\label{tab:exp_finetune_chatbot_instruct}
\end{table}

\subsection{Podcast Generation Evaluation}
To evaluate zero-shot podcast generation, we curated two two-speaker podcast evaluation sets: dialogue-zh and dialogue-en, containing 100 Mandarin and 115 English dialogues, respectively. Each dialogue test set spans 4 to 10 turns, totaling 1.67 hours for Mandarin and 2.35 hours for English. For each dialogue, we use the first two turns as the prompt and generate the remaining turns with the post-trained FireRedTTS-2 model.

We assess the generated dialogues using three objective metrics: intelligibility, speaker similarity, and Mel-cepstral distortion (MCD)\footnote{\url{https://github.com/chenqi008/pymcd}}. For intelligibility, we use Whisper-large-v3\footnote{\url{https://huggingface.co/openai/whisper-large-v3}} to compute word error rate (WER) for English and Paraformer-zh\footnote{\url{https://huggingface.co/funasr/paraformer-zh}} to compute character error rate (CER) for Mandarin. Speaker similarity is measured with the WavLM-Large model\footnote{\url{https://github.com/microsoft/UniSpeech/tree/main/downstreams/speaker_verification}}. For subjective evaluation, we conduct a comparative mean opinion score (CMOS) test in which raters choose the more natural synthesized dialogue between FireRedTTS-2 and competing models. We compare against open-source dialogue TTS systems including MoonCast, ZipVoice-Dialog, and MOSS-TTSD\footnote{We were unable to generate the test set with Dia and Sesame due to instability and therefore exclude them from comparison.}. Because MoonCast, ZipVoice-Dialog, and MOSS-TTSD produce a single mixed-track dialogue containing both voices, we use Pyannote\cite{Plaquet23pyannote1, Bredin23pyannote2} to segment each speaker before evaluation. The results are listed in Table \ref{tab:exp_zeroshot_dialogue}.

\begin{table}[htp]
\centering
\resizebox{\columnwidth}{!}{
\begin{tabular}{c|cccc|cccc}
\toprule
\multirow{2}{*}{\textbf{Model}} & \multicolumn{4}{c|}{\textbf{dialogue-zh}} & \multicolumn{4}{c}{\textbf{dialogue-en}} \\ \cmidrule{2-9} 
 & \textbf{CER$\downarrow$} & \textbf{SIM$\uparrow$} & \textbf{MCD$\downarrow$} & \textbf{CMOS$\uparrow$} & \textbf{WER$\downarrow$} & \textbf{SIM$\uparrow$} & \textbf{MCD$\downarrow$} & \textbf{CMOS$\uparrow$} \\ \midrule
MoonCast & 3.81 & 0.658 & 11.37 & -0.21 & 3.81 & 0.620 & 10.96 & -0.21 \\
ZipVoice-Dialog & 2.93 & 0.736 & 9.29 & -0.18 & 11.71 & 0.701 & 9.88 & -0.31 \\
MOSS-TTSD & 3.99 & 0.659 & 8.32 & -0.16 & 5.43 & 0.550 & 9.25 & -0.13 \\ \midrule
FireRedTTS-2 & \textbf{2.08} & \textbf{0.753} & \textbf{7.99} & \textbf{0.0} & \textbf{3.16} & \textbf{0.703} & \textbf{9.06} & \textbf{0.0} \\ \bottomrule
\end{tabular}
}

\caption{Objective and subjective evaluation of zero-shot podcast generation.}
\label{tab:exp_zeroshot_dialogue}
\end{table}

Table \ref{tab:exp_zeroshot_dialogue} shows that FireRedTTS-2 delivers the most stable synthesis, achieving the lowest WER/CER on dialogue‑zh and dialogue‑en, suggesting that our lower frame rate and semantically enhanced speech tokenizer enable robust modeling of long speech sequences. It also attains the highest speaker similarity, reflecting strong cross‑turn voice cloning and reliable speaker transitions, and the lowest MCD, indicating minimal deviation from ground truth; CMOS results further confirm its contextually coherent naturalness and highlight the effectiveness of the dual‑transformer’s context‑learning capabilities.

\begin{figure}[htp]
\centering
\includegraphics[width=0.65\linewidth]{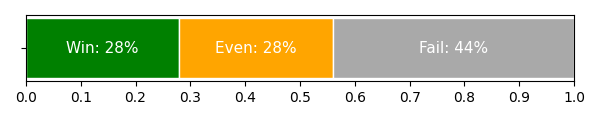}
\caption{Subjective preference results between FireRedTTS-2 fine-tuned on two podcast speakers and ground truth recordings. "Win": FireRedTTS-2 synthesis is more natural than ground truth dialogue speech; "Even": indistinguishable; "Fail": ground truth is more natural.}
\label{img:exp_feihua_cmos}
\end{figure}

We also evaluate the intelligibility and naturalness of the FireRedTTS-2 model fine-tuned on two podcast speakers using the dialogue-zh set. For intelligibility, the fine-tuned model maintains stable synthesis performance and achieves a lower CER of 1.66\% than in zero-shot mode. For naturalness, we conduct a subjective test in which raters are asked to choose the more natural sample between the FireRedTTS-2 synthesis and the ground truth dialogue. As shown in Figure \ref{img:exp_feihua_cmos}, our synthesis is preferred as more natural in 28\% of cases and judged equally natural in another 28\%, matching or surpassing real recordings in 56\% of trials and indicating that FireRedTTS‑2 can produce human‑like podcast speech.


\section{Conclusions}
In this work, we present FireRedTTS-2, a conversational TTS system for dialogue-centric applications such as interactive chat and podcast generation. It comprises a newly developed speech tokenizer and a text-to-speech model. The tokenizer operates at a low 12.5Hz frame rate and encodes richer semantics, shortening speech sequences and facilitating robust text-to-token modeling; it also supports high-fidelity streaming decoding, making the system well suited to real-time use. The text-to-speech model adopts a dual-transformer architecture and a text–speech interleaved format to enable flexible sentence-by-sentence generation with first-packet latency under 100 ms, serving both real-time interaction and offline podcast production. Experiments show that, in monologue settings, FireRedTTS-2 offers strong zero-shot voice cloning comparable to conventional monologue TTS systems; in dialogue settings, it can be integrated seamlessly into existing frameworks without adjusting other modules and produces emotionally expressive speech inferred from implicit contextual cues. In zero-shot podcast generation, it delivers more stable synthesis, more accurate speaker transitions, and more coherent prosody, outperforming state-of-the-art dialogue generation models. Furthermore, it can be tailored to two specific podcast voices, producing dialogue speech that is difficult to distinguish from human recordings.

\bibliographystyle{unsrt}
\bibliography{refs}

\end{document}